\documentclass[final, runningheads]{llncs}
\usepackage[T1]{fontenc}
\usepackage{graphicx}
\usepackage{subfig}
\usepackage{amsmath}
\newcommand{\itadata}{\footnotesize \textsl{Workshop Scientific HPC in the pre-Exascale era (part of ITADATA2024)}}
\usepackage{fancyhdr}
\fancypagestyle{empty}{\fancyhf{}\fancyhead[C]{\itadata}
  \fancyhead[R]{\includegraphics[width=.05\textwidth]{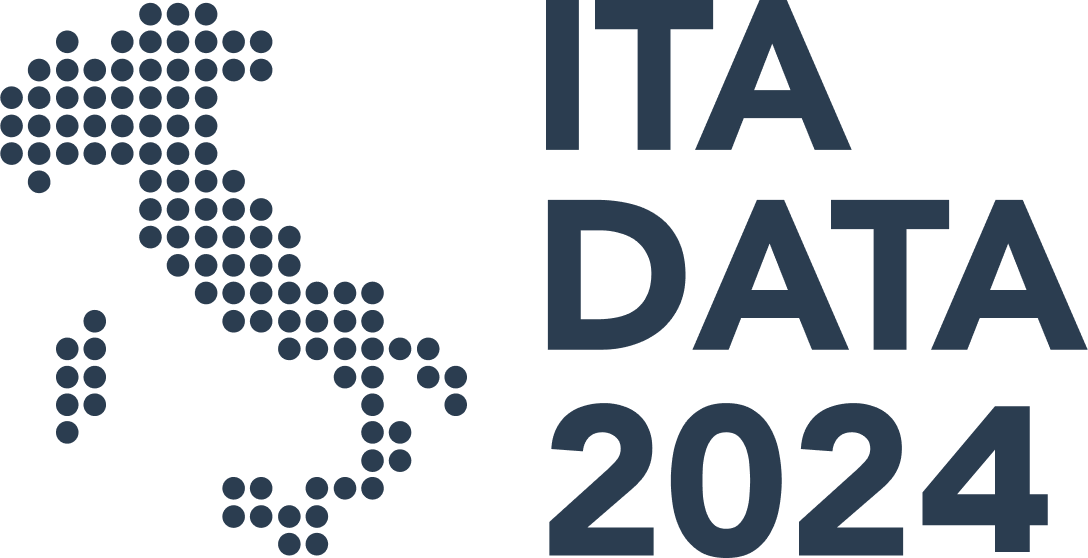}}}

\usepackage{xcolor}

\makeatletter
\begin{document}
\title{Exploring energy consumption of AI frameworks on a 64-core RV64 Server CPU}
\author{Giulio Malenza\inst{1}\orcidID{0009-0006-4862-7429} \and
Francesco Targa\inst{1}\orcidID{0009-0000-6498-2179} \and
Adriano Marques Garcia\inst{1}\orcidID{ 0000-0003-4796-773X} \and
Marco Aldinucci\inst{1}\orcidID{0000-0001-8788-0829} \and
Robert Birke\inst{1}\orcidID{0000-0003-1144-3707}}
\authorrunning{F. Author et al.}
\institute{University of Torino, Italy.}
\maketitle              \begin{abstract}
In today's era of rapid technological advancement, artificial intelligence (AI) applications require large-scale, high-performance, and data-intensive computations, leading to significant energy demands. Addressing this challenge necessitates a combined approach involving both hardware and software innovations. Hardware manufacturers are developing new, efficient, and specialized solutions, with the RISC-V architecture emerging as a prominent player due to its open, extensible, and energy-efficient instruction set architecture (ISA). Simultaneously, software developers are creating new algorithms and frameworks,
yet their energy efficiency often remains unclear. 
In this study, we conduct a comprehensive benchmark analysis of machine learning (ML) applications on the 64-core SOPHON SG2042 RISC-V architecture. We specifically analyze the energy consumption of deep learning inference models across three leading AI frameworks: PyTorch, ONNX Runtime, and TensorFlow. Our findings show that frameworks using the XNNPACK back-end, such as ONNX Runtime and TensorFlow, consume less energy compared to PyTorch, which is compiled with the native OpenBLAS back-end.


\end{abstract}
\section{Introduction}

In recent years, artificial intelligence (AI) has driven significant advancements across multiple sectors, from healthcare to finance and beyond. However, these AI applications, particularly those involving deep learning, require substantial computational resources, leading to increased energy consumption. As global concerns about energy efficiency and sustainability grow, exploring methods to optimize the energy use of these resource-intensive systems is critical.

The RISC-V architecture~\cite{pattersonbook} has emerged as a key player in this landscape, offering an open, extensible instruction set architecture (ISA) that emphasizes energy efficiency. It has made RISC-V architectures promising candidates for developing and deploying more sustainable AI solutions. Meanwhile, various AI frameworks, such as PyTorch, TensorFlow, and ONNX Runtime, have been designed to harness the computational power of modern hardware. However, the energy consumption of these frameworks, particularly on high-performance multicore RISC-V systems, remains underexplored.

This paper presents a comprehensive exploration analysis of AI inference workloads on a 64-core SOPHON SG2042 RISC-V system. Examining popular AI frameworks' energy consumption and performance across different deep learning models is a significant step towards providing insights into how hardware and software optimizations can contribute to more energy-efficient AI solutions. 

The main contributions of this work can be summarized as follows:
\begin{itemize}
    \item We installed three of the most commonly used AI frameworks --PyTorch, TensorFlow Lite, and ONNX Runtime-- on RISC-V architecture.
    \item We verified the correct predictions using three pretrained models having different architectures --ResNet-50, VGG-16, and MobileNetV2.
    \item We experimentally measured the time- and energy-to-solution of the considered AI frameworks.
\end{itemize}

\paragraph{Paper roadmap}: Section \ref{sec2} introduces literature works related to our analysis. Section \ref{sec3} presents the specifications of the devices used in this work. Section \ref{sec4} describes the AI framework analyzed, and the inference models used. Section \ref{sec5} discusses the results obtained. Section \ref{sec6} concludes the paper by highlighting possible future works.

\section{Related Work}\label{sec2}

Regarding energy consumption of deep neural networks deployed on RISC-V platforms, Nagar et al.~\cite{Nagar2023} present a study on accelerating facial recognition using the convolutional neural network (CNN)-based MobileNet model on embedded systems, including a RISC-V-based processor. They benchmark a facial recognition application on the Sipeed Maixduino, a RISC-V development board equipped with a Knowledge Processing Unit (KPU) accelerator. Their results indicate a substantial reduction in power consumption and execution time compared to traditional CPU implementations, underscoring the efficacy of RISC-V architectures for energy-constrained applications. While this study focuses on a specific use case with a single deep neural network (DNN) model and a small dual-core RISC-V architecture, our work extends this evaluation by testing multiple DNN frameworks, including PyTorch, TensorFlow, and the ONNX runtime, on a more powerful 64-core RISC-V CPU. 

Suárez et al.~\cite{suarez2024} conduct a comprehensive evaluation of energy efficiency and performance across ARM and RISC-V System-on-Chips (SoCs), providing a detailed comparative analysis using the NAS Parallel Benchmarks and TensorFlow Lite (TFLite) benchmarks. Their study examines power consumption, execution time, and energy efficiency on different hardware platforms, including an in-depth case study using Computational Fluid Dynamics (CFD) simulations with OpenFOAM. The findings suggest that while RISC-V SoCs may demonstrate lower average power consumption, they do not consistently outperform ARM counterparts in terms of performance per watt. 

Mittone et al.~\cite{Mittone2023} delve into the use of emerging RISC-V systems for decentralized machine learning (DML), explicitly targeting Federated Learning and Edge Inference across various hardware architectures, including a SiFive quad-core RISC-V CPU. A key contribution of their work is the introduction of the first publicly available RISC-V port of the PyTorch framework, which allows for evaluating performance and energy efficiency in distributed machine learning tasks. Their study highlights the potential of RISC-V for energy-efficient DML, particularly in scenarios that require flexible and decentralized learning schemes.

Lee et al.~\cite{Lee2023} introduce a novel approach to the efficient DNN training on RISC-V architectures through Dynamic Block Size and Precision Scaling (DBPS). Their proposed technique dynamically adjusts block sizes and precision during training to minimize energy consumption while maintaining training accuracy. The DBPS method is integrated into a RISC-V processor via custom instruction set architecture (ISA) extensions and is supported by a specialized hardware accelerator. Evaluating their system on an FPGA prototype with various benchmarks, including ResNet-18 on CIFAR-10 and ImageNet, the authors report a reduction in training time and energy consumption by 67.1\% and 72.0\%, respectively. This work highlights the potential of RISC-V ISA extensions for DNN applications.

While previous works have explored the energy efficiency of DNNs on RISC-V architectures, including specialized use cases like facial recognition and decentralized machine learning, our research goes beyond these studies by evaluating multiple deep learning frameworks—such as PyTorch, TensorFlow, and ONNX runtime on a 64-core multicore RISC-V CPU. This comprehensive approach offers more profound insights into the scalability and performance of these frameworks on advanced RISC-V hardware, contributing to a better understanding of DNN energy consumption across different runtimes and hardware configurations.

\section{Methodology} \label{sec3}
\subsection{SOPHON SG2042 SoC}
The SG2042 system-on-chip (SoC) contains 64 RISC-V cores divided into 16 clusters connected through a grid network. Each cluster comprises a XuanTie C920 4-core RISC-V CPU.
Each core has 64KB of L1 instruction and data cache, and each cluster of 4 cores shares a 1MB of L2 cache. 
The unified L2 cache can handle two access requests in parallel within one cycle.
The grid interconnect finally offers access to 64MB of level 3 cache shared among all 64 cores. 
Four DDR4-3200 memory controllers manage access to the main memory system.
For peripherals, the SG2042 is equipped with 32 PCIe Gen4 lanes.

The XuanTie C920~\cite{Xuantie-910} is a homogeneous high-performance 64-bit multi-core RISC-V CPU architecture designed by T-Head that supports 1 to 4 cores at a maximum operation frequency of 2 GHz. 
It targets high-performance applications and implements a 12-stage, out-of-order, multiple-issue superscalar pipeline.
Based on the RISC-V instruction set architecture (ISA), this CPU provides the RV64GCV instruction set~\cite{RV-ISA_manual:2014}, which supports standard vector instructions extension version 0.7.1 (RVV 0.7.1).
The vector processing unit's vector registers are 128 bits long and support the FP16, FP32, FP64, INT8, INT16, INT32, and INT64 data types.

Milk-V Pioneer Box is a commercial and ready-to-use development platform in the form of a desktop computer. It is equipped with a Pioneer motherboard powered by the SOPHON SG2042.
It was designed to provide a native development environment for developers and researchers.
It eliminates or reduces the need to use different platforms to cross-compile programs for the RISC-V architecture. 
The one we use in this paper has 128GB of RAM DDR4 (3200MHz) and a 1TB PCIe 3.0 SSD.
The operating system is Linux fedora-riscv 6.1.31.

\subsection{TAPO P125M}
SOPHON SG2042 has no hardware counters dedicated to energy/power consumption. Hence, as done by~\cite{riscvhpcenergy}, we used the TP-Link Tapo P125M smart plug to measure power and energy consumption. Since there is no official library to enable remote monitoring with scripts, we found a nonofficial version in the GitHub repository\footnote{https://github.com/mihai-dinculescu/tapo/tree/main}. After checking the correctness of the measurements, we used the library to measure the power, in watts (W), and energy consumption, in watt-hours (Wh), of our simulations. The library can be installed with a standard \texttt{pip} installer and allows monitoring both the instantaneous power, via \texttt{get\_current\_power()} and consumed energy, via \texttt{get\_energy\_usage()}. Based on these two functions, we created a Python script that starts the simulation and reads out power and energy values at regular intervals. Although these two functions are computationally lightweight, it is reasonable to think that the measurements introduce compute and, hence, energy overhead.

By default, the tapo smart plug can take measurements at each second. This limits the simulation time to be greater than one second. However, reading the values every second could result in excessive overhead. Here we chose as sampling time 10s to accommodate the runtimes of the shorter experiments as well as limit the overhead.

\section{Evaluation}\label{sec4}

\subsection{AI Frameworks}
In this preliminary study, our primary objective is to measure the energy consumption of various AI frameworks when performing classical inference tasks. Given the current limitations of the software stack on the emerging RISC-V architecture, we explored which frameworks could be fully or partially installed on the system. We successfully installed three frameworks: PyTorch, TensorFlow Lite, and ONNX Runtime. For each, we executed inference models to assess their accuracy. All the frameworks were compiled with the Xuantie-Gcc compiler in release mode \cite{xuantie-gcc}.

PyTorch~\cite{Torch} is one of the most famous and used deep learning frameworks. It is designed to be user-friendly and Pythonic, making it easy for researchers and developers to write models, data loaders, and optimizers. Nonetheless, PyTorch is closely related to the Python environment, as performance is crucial to training and testing ever-larger deep learning models; most of it is written in C++. Moreover, it can delegate the core computations to specialized high-performance linear algebra libraries to fully exploit all the hardware capabilities; in particular, we used OpenBLAS library version 0.3.26 without support for the RISC-V vector extensions for fairness to the other frameworks.

The first porting to RISC-V was done by Colonnelli et al.~\cite{23:risc-v-summit}. Specifically, they ported the Chromium Breakpad library, the SLEEF vector math library, and the PyTorch CPU Information library (cpuinfo) to RISC-V. This porting enabled the authors to compile and test PyTorch v2.0 on the SiFive Freedom U740 SoC machine. They tested applications included in the default PyTorch benchmark suite. For training, they selected the regression and MNIST applications, while for inference, they used ImageNet with various image classification models.

The second framework analyzed is Tensorflow~\cite{Tensorflow}, a machine learning system that strongly supports deep learning networks' training and inference, made by Google. TensorFlow is designed to be scalable, making it suitable for both research and production environments, from small devices to large-scale distributed systems. A lightweight framework version, TensorFlow Lite~\cite{Tflite-RISCV} (Tflite), was introduced to optimize inference problems on mobile and embedded devices~\cite{spcl}. This is because inference model applications usually do not require the same computational power as training the models but can require frequent data control and can be executed on small devices. The framework was developed to support domain-specific ISA extensions for machine learning; as the target ISA, the authors used RISC-V ISA Vector extensions~\cite{Patterson}. As in the previous case, the framework was developed to use high-performance linear algebra libraries. We could not compile TensorFlow from scratch due to lack of RISC-V Java support for the Bazel build system. We compiled the Lite version of the framework with XNNPACK~\cite{XNNPACK-RV}, a Google library developed to optimize neural network computations across different hardware platforms. XNNPACK can also be compiled using the RISC-V vector extension. However, we did not enable it (\texttt{XNNPACK\_ENABLE\_RISCV\_VECTOR=OFF}) because it requires RVV 1.0, which is unavailable on our machine. We need XNNPACK because, in our tests, Tflite tries to use it, returning an error if it is not found.

The last framework tested is ONNX Runtime~\cite{onnxruntime}. The library is specialized in optimizing and executing machine learning inference ONNX models. It is designed to be flexible and capable of executing inferencing in many different languages and hardware stacks.
The library exposes different execution providers to optimally process the ONNX models on the hardware platform. We tested default CPU providers but noticed that the predictions were wrong; compiling the framework enabling the XNNPACK execution provider solved the problem. 

\subsection{AI Inference models}
We chose three classical deep learning models to test the power and energy consumption of the previously introduced AI frameworks.
The first model is VGG-16~\cite{VGG16}, a straightforward convolutional neural network architecture proposed by Oxford University in 2014 during the year's ImageNet Large-Scale Visual Recognition Challenge (ILSVRC). The main difference with previous networks~\cite{prev1,prev2} relies on the smallest receptive fields used during the convolutional layers. Increasing the convolutional linear layers while keeping the number of receptive fields small allows for improving both accuracy and performance with respect to previous networks.

Introduced in 2015 at the ILSVRC  challenge, ResNet~\cite{Resnet} represented a crucial step toward the solution of the well-known degradation problem~\cite{deg1,deg2} arising when the network depth increases, in particular, increasing the number of layers also increases the training and testing errors. The paper introduces a deep residual learning framework that allows better gradient propagation by skipping some layers using ``shortcut connections''~\cite{shortcut1,shortcut2}. The skip connections limit the degradation problem, allowing the stack of any number of blocks for a network with arbitrary depth.

The last tested model is MobileNet~\cite{MobileNet}. This model was introduced by Google in 2017 as a class of models for mobile and embedded applications. It is based on depthwise separable convolutions~\cite{mob1,mob2}, which allows to achieve high accuracy with low computational costs. Despite its efficiency, MobileNet maintains competitive accuracy, making it a popular choice for deploying deep learning models on mobile and edge devices.

\subsection{Results}\label{sec5}

The power consumption analysis of the AI frameworks was conducted using a subset of the Imagenet dataset~\cite{imagenetdataset} as the test data. This selection was made because the pre-trained models were initially trained on this dataset, allowing us to both assess inference accuracy and verify the proper execution of the frameworks. Although the full dataset contains 3,923 images, we found that a representative subset of 1000 images was sufficient for our analysis. Therefore, a subset of 1000 images was used in this study. 
Table~\ref{tab:model_comparison} shows the accuracy obtained from varying models and frameworks on this sub-dataset. Given the high values achieved, we are confident that the frameworks are properly built and the inference models work correctly.
\begin{table}[h]
\centering
\begin{tabular}{|c|c|c|c|}
\hline
\textbf{}        & \textbf{ResNet-50} & \textbf{VGG-16} & \textbf{MobileNetV2} \\ \hline
\textbf{TFLite}  & 83.7\%            & 76.9\%         & 76.6\%               \\ \hline
\textbf{Torch}   & 83.7\%            & 78.0\%         & 76.6\%               \\ \hline
\textbf{ONNX}    & 81.6\%            & 76.5\%         & 76.6\%               \\ \hline
\end{tabular}
\caption{Comparison of model accuracy across different frameworks}
\label{tab:model_comparison}
\end{table}

\begin{figure}
    \centering
    \includegraphics[width=0.7\textwidth]{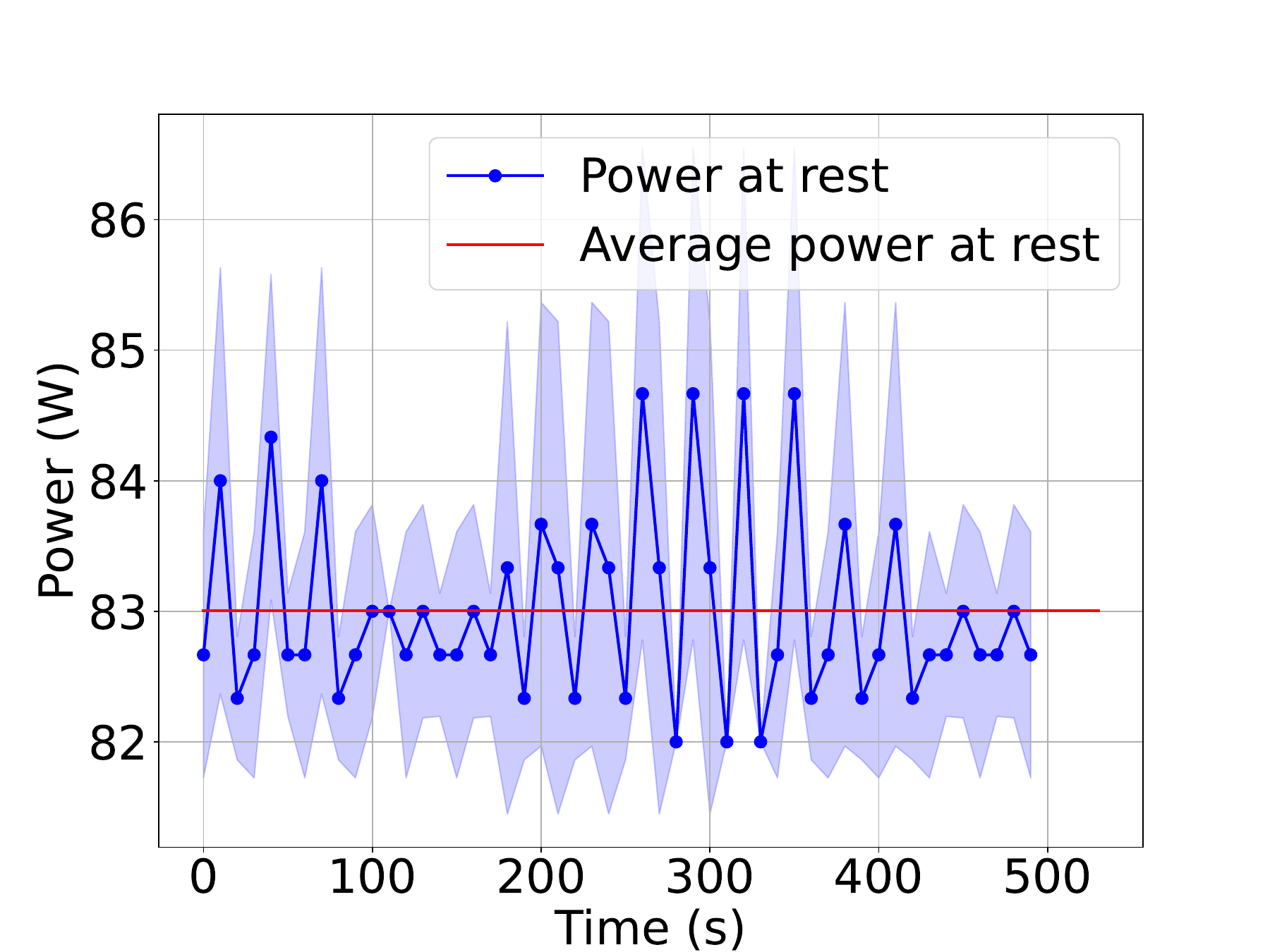}
    \caption{Power at rest consumed by the device.}
    \label{fig:PowerRest}
\end{figure}

\begin{figure*}[t!]
   \subfloat[ResNet-50\label{fig:Res50_scal}]{
      \includegraphics[width=0.33\textwidth]{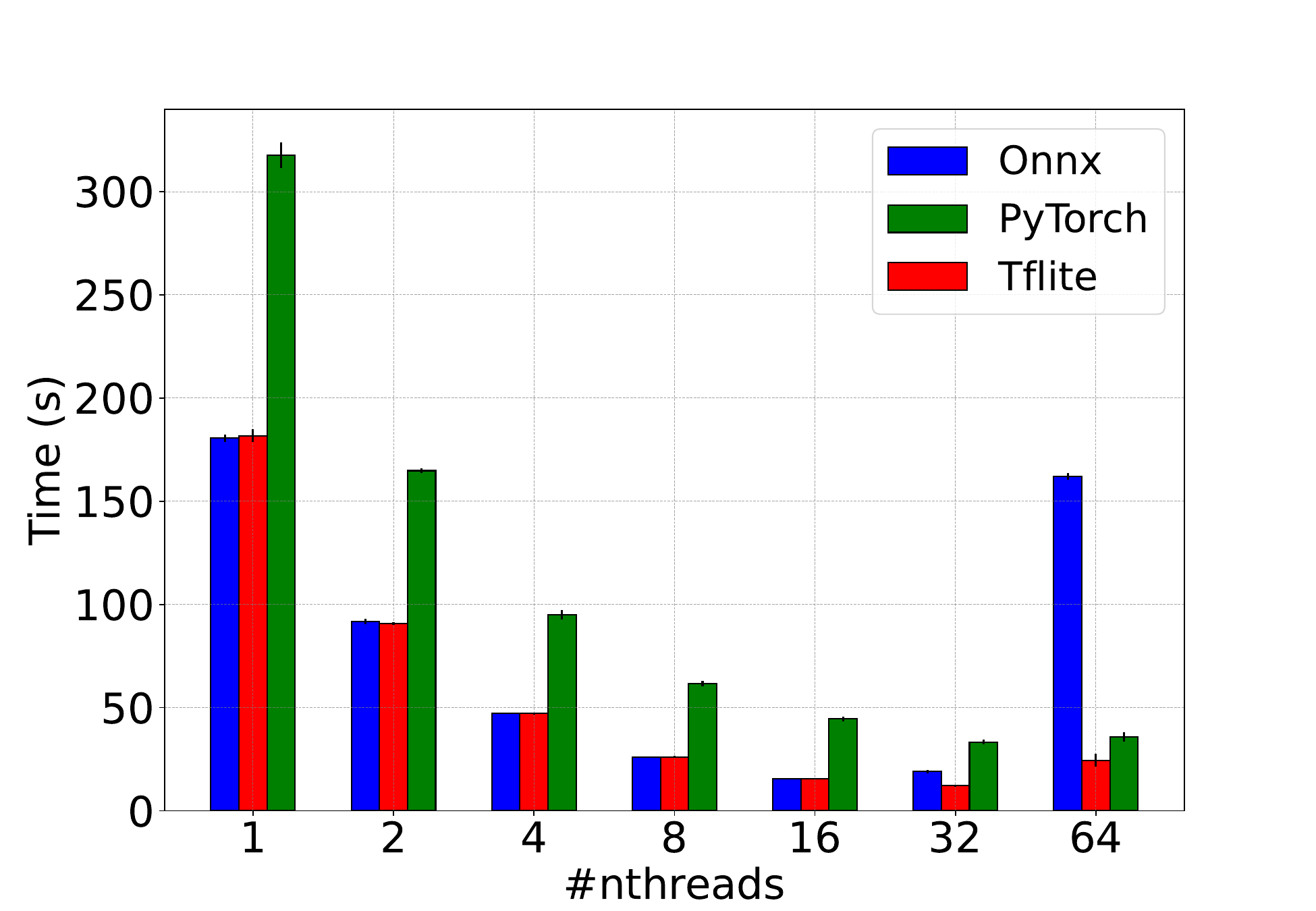}}
   \subfloat[VGG-16\label{fig:vgg16_scal}]{
      \includegraphics[width=0.33\textwidth]{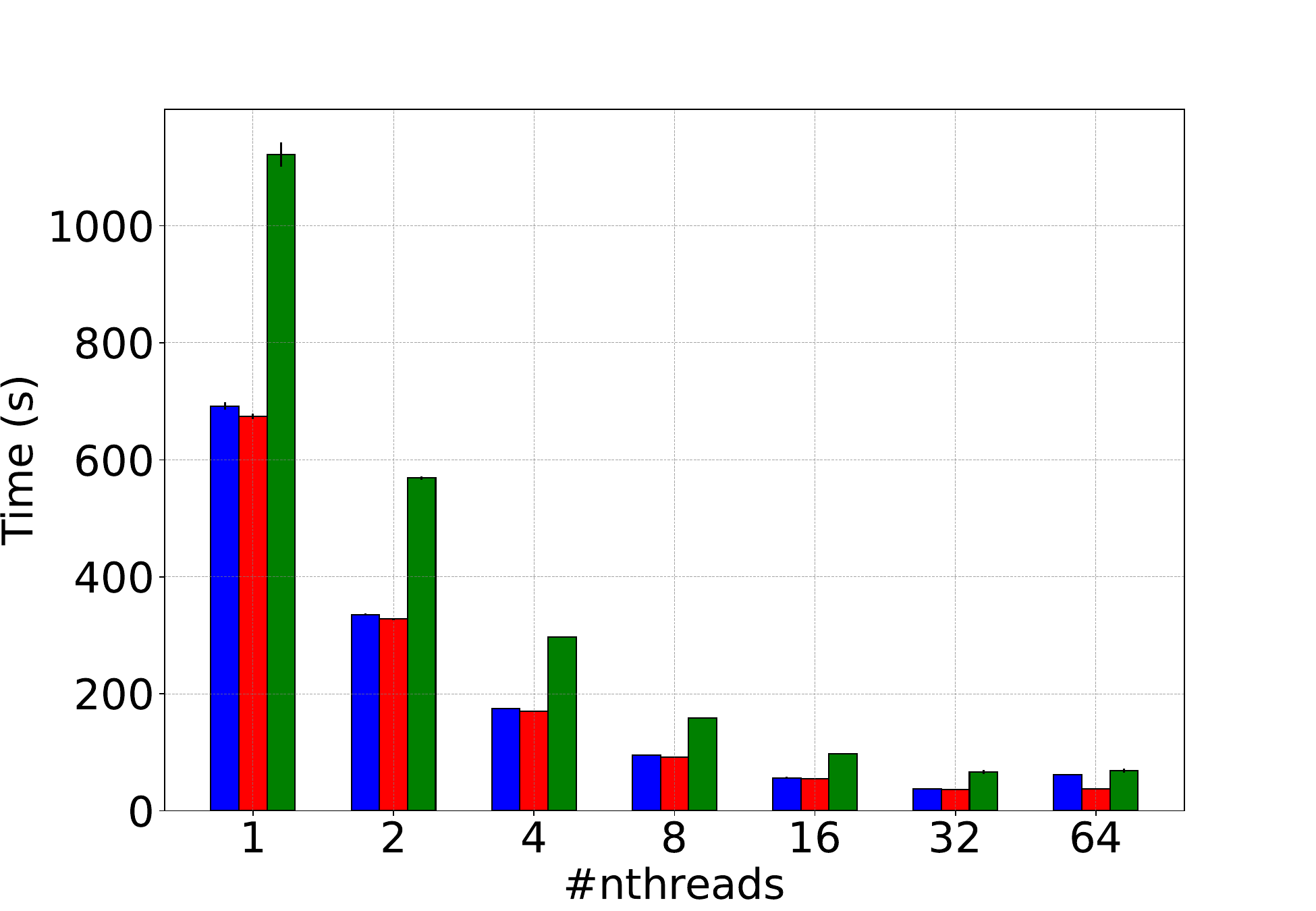}}
   \subfloat[MobilenetV2\label{fig:MobV2_scal}]{
      \includegraphics[width=0.33\textwidth]{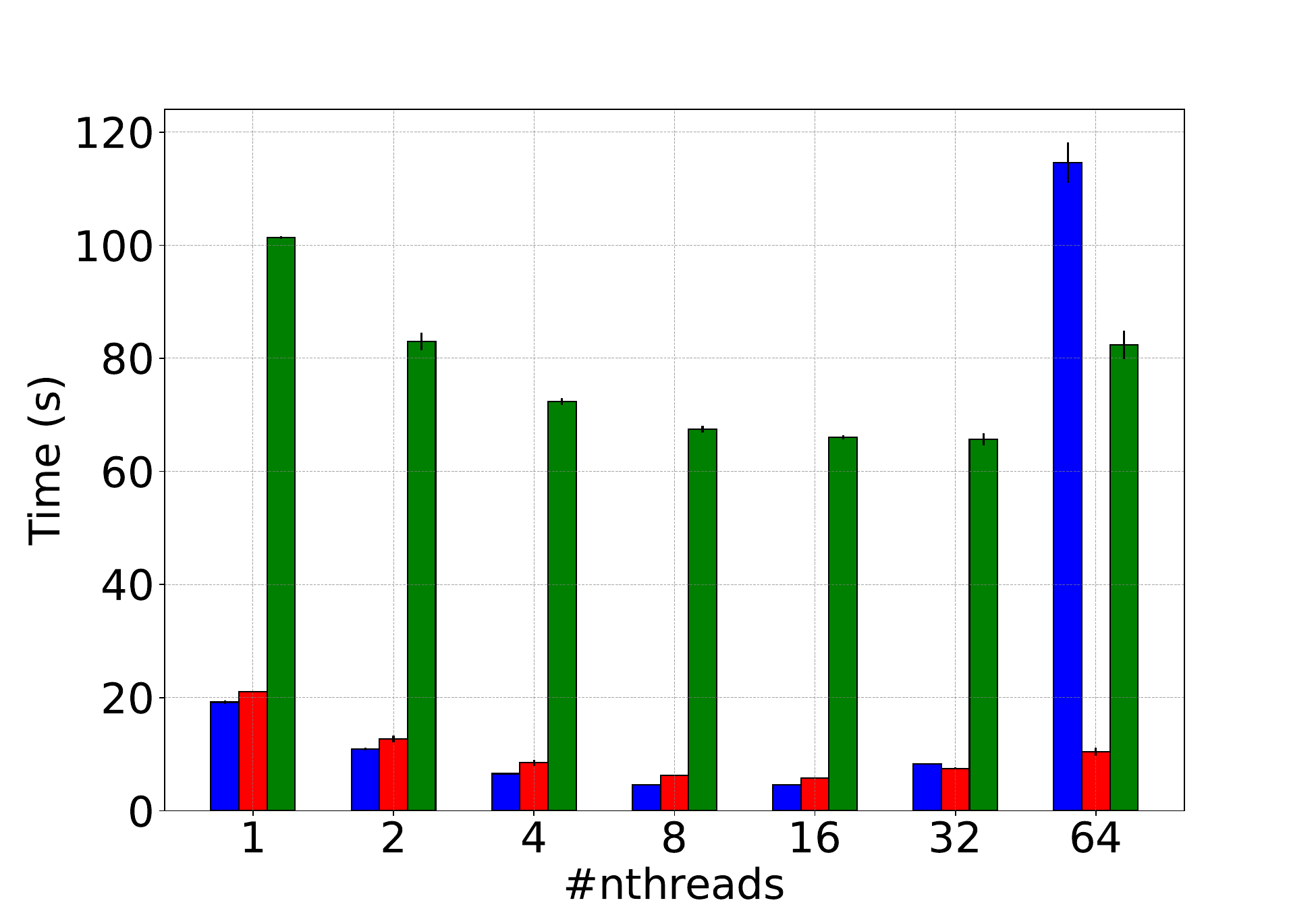}}\\
\caption{Resnet-50, VGG-16 and MobileNetV2 inference scaling simulations.}
    \label{fig:scaling}
\end{figure*}

Before measuring the frameworks' actual power and energy consumption, we estimated the power and energy consumed by the device at rest when it is turned on and only the operating system is executing. Figure~\ref{fig:PowerRest} shows the average power consumed by the device at rest. We repeated the test three times to compute the average and standard deviation error.
Energy and power values are read at time intervals of 10s for a total simulation time of 500s, consistent with the inference simulations. On average, the power consumed by the device at rest is $83.01\pm 0.99$ W (the red line in the figure). In the period considered, the average energy consumed is $11.33\pm 0.47$ Wh.  
The AI frameworks analyzed were developed to enable different layers of parallelism, exploiting both multicore and GPU hardware capabilities. In this work, we are considering only multicore parallelism. Ideally, increasing the number of threads improves performance. However, this depends on how much the framework and libraries it uses are optimized and how the architecture was designed. Exploiting full multi-threaded parallelism can generate NUMA and memory bandwidth effects, leading to performance degradation. For this reason, before running simulations to calculate energy consumption, we performed inference tests by varying the number of threads used by the frameworks. Figures~\ref{fig:scaling} show the results of our tuning analysis. Simulations were performed on a dataset of 100 images; they were repeated ten times to take the average time with its corresponding standard deviation error. For the Resnet-50 model, the best time performance was achieved using TensorFlow Lite $12.21\pm 0.49$ (s) with 32 cores. ONNX Runtime processes 100 images in $15.43\pm 0.31$ (s) using 16 cores, while PyTorch employs $33.27\pm 1.15$ (s) with 32 cores. VGG-16 model was processed by ONNX Runtime in $37.76\pm 0.47$ (s) using 32 cores. With the same number of cores, TensorFlow Lite and Pytorch employ $36.31\pm 0.42$ and $66.12\pm 3.80$ seconds, respectively. In contrast to the other models, execution on 64 cores did not exhibit significant overhead. This is likely due to the model's larger size, which allows it to effectively utilize all 64 cores, and because its operations are less computationally intensive compared to the other models. The last model, MobileNetV2, was executed by ONNX Runtime in $4.55 \pm 0.11$ (s) using 16 cores. With the same number of cores, TensorFlow Lite employs $5.835\pm 0.228$ (s) while PyTorch takes $65.70\pm 1.22$ (s) with 32 cores. 

As expected, increasing the number of cores improves performance until reaching the minimum ($<64$). After that, performance slightly degradate. This is a common issue in multicore systems, where the overhead of coordinating multiple cores takes longer than executing computational operations.

\begin{figure*}[t!]
   \subfloat[ResNet-50\label{fig:Res50}]{
      \includegraphics[width=0.32\textwidth]{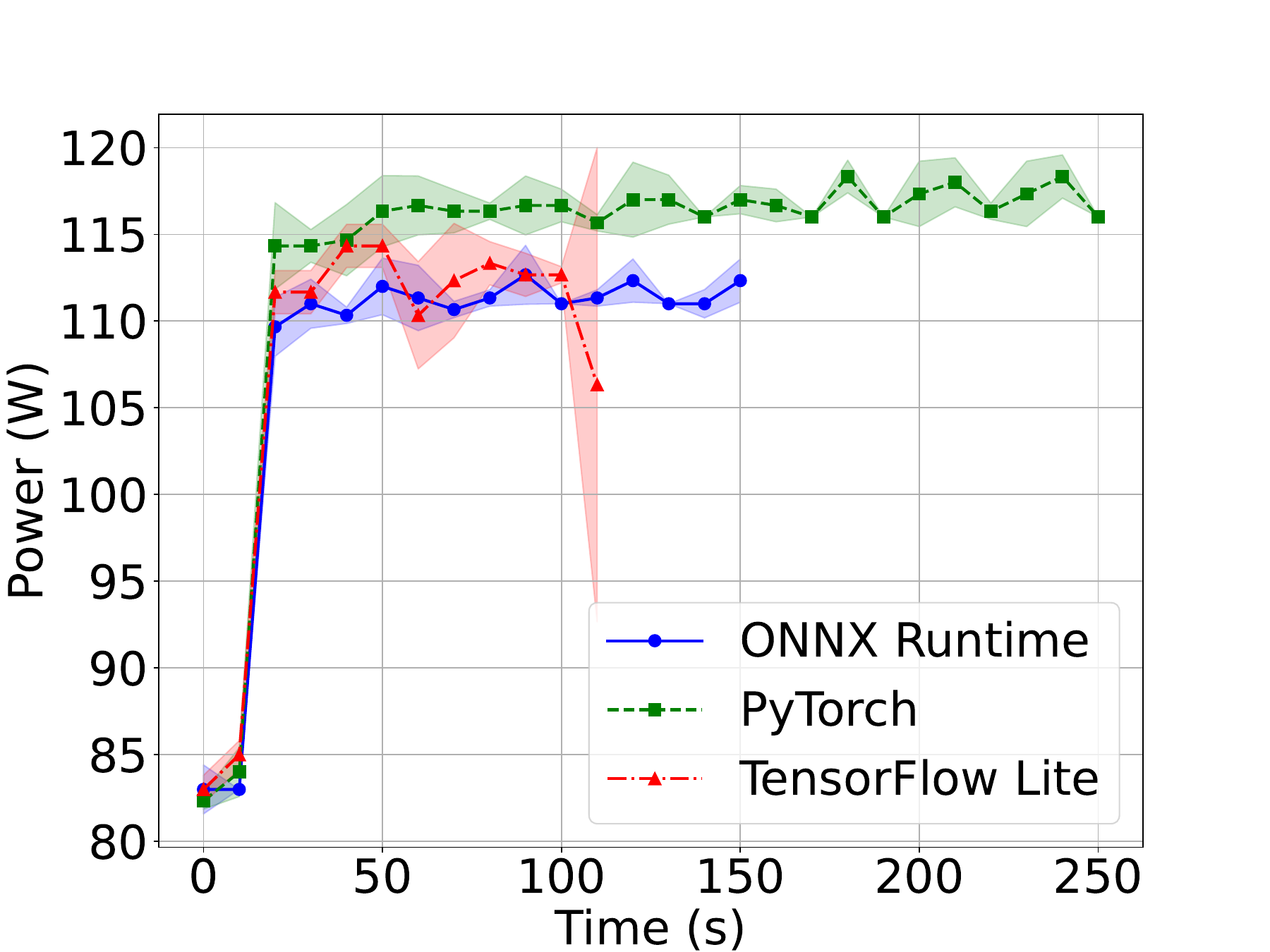}}
\hspace{\fill}
   \subfloat[VGG-16\label{fig:vgg16} ]{
      \includegraphics[width=0.32\textwidth]{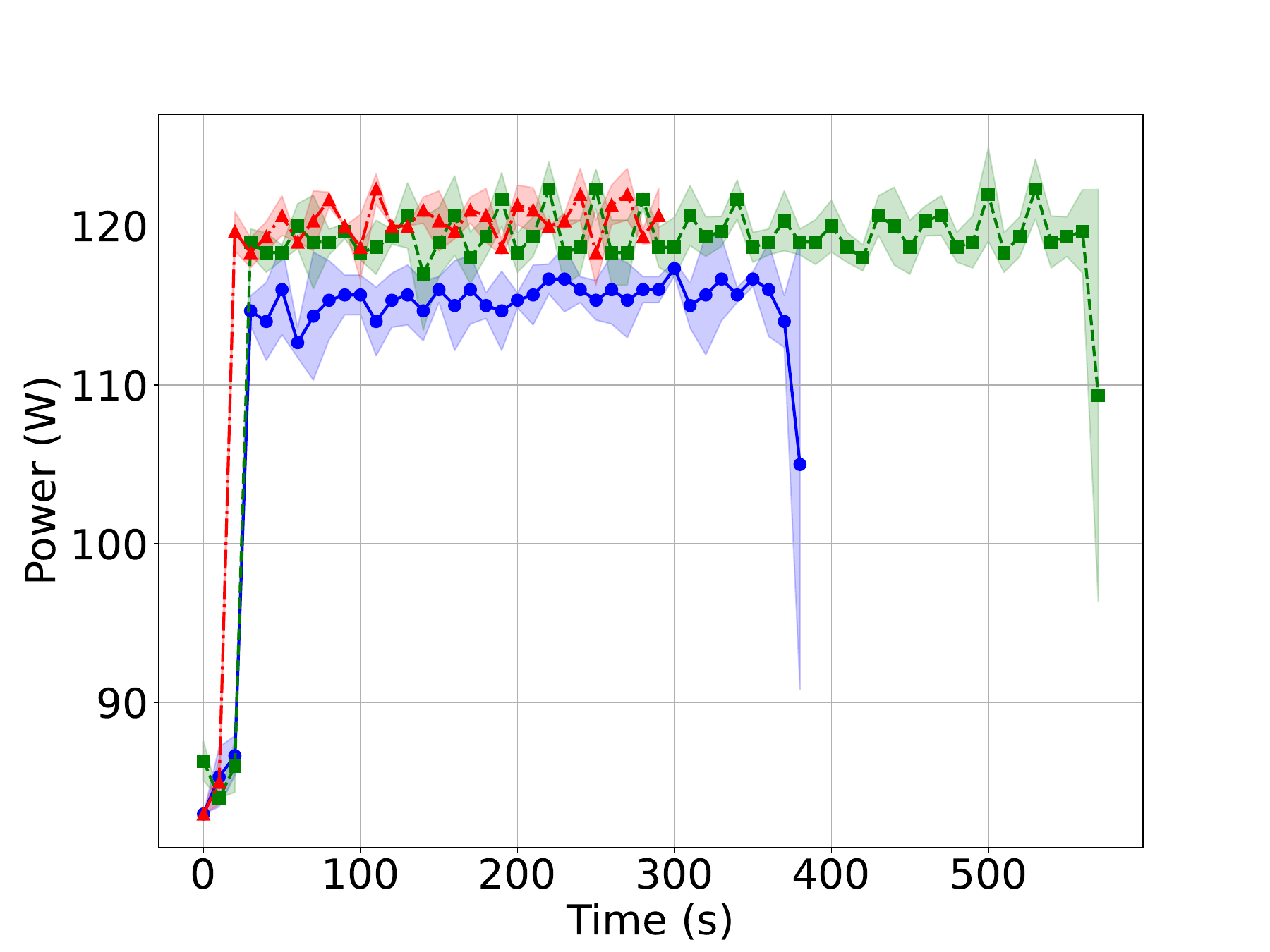}}
\hspace{\fill}
   \subfloat[MobileNetV2\label{fig:MobV2}]{
      \includegraphics[width=0.32\textwidth]{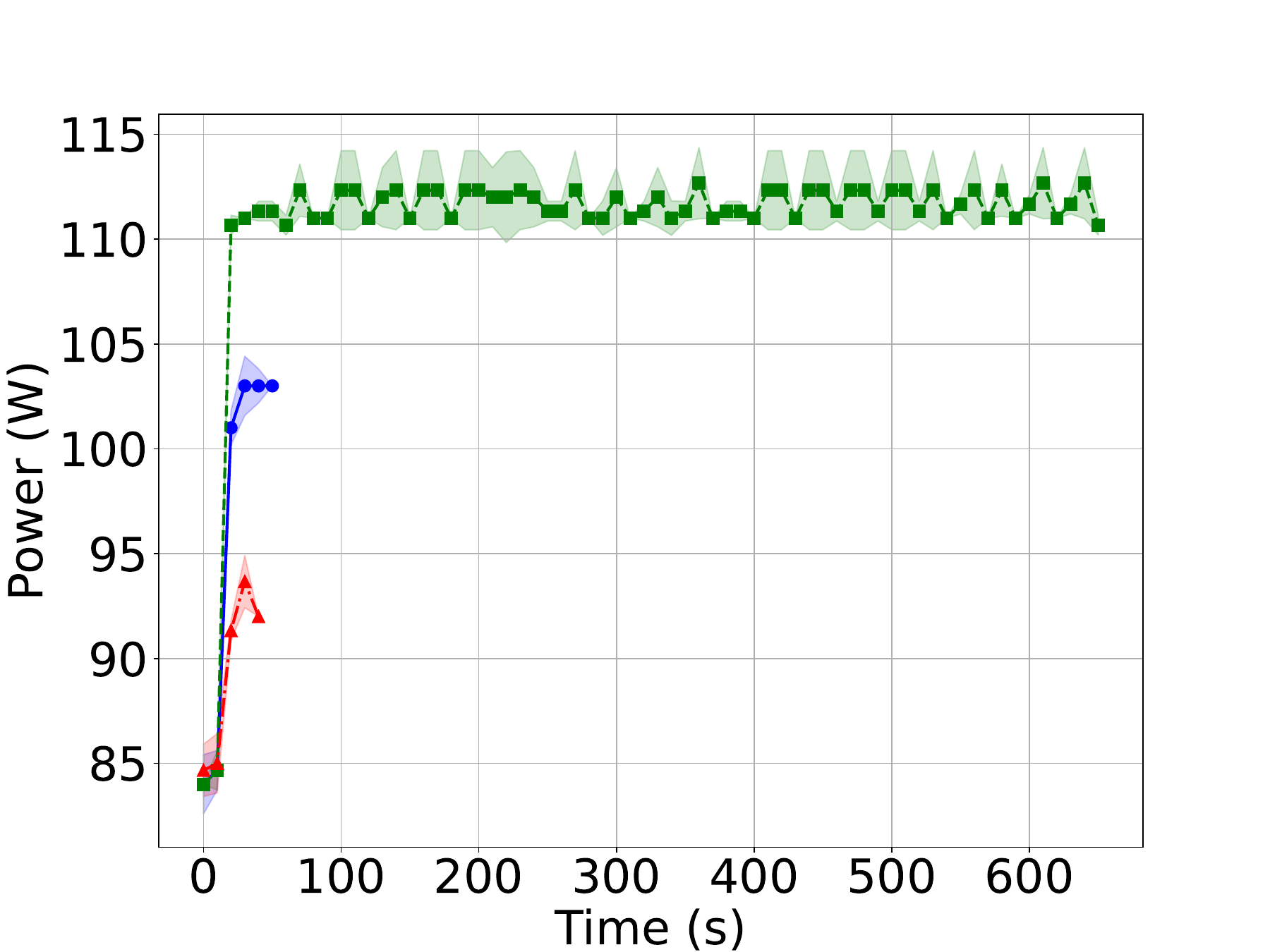}}\\
\caption{ResNet-50, VGG-16 and MobileNetV2 inference power consumption, simulations on 1000 images.}
    \label{fig:sys}
\end{figure*}

Figure~\ref{fig:sys} shows the power consumed during the inference model execution varying the frameworks, configured on the basis of the previous analysis. Figure~\ref{fig:Res50} represents the ResNet-50 model. In this case, the best time performance was obtained using TensorFlow lite. Its inference total execution time took $124.19\pm 4.79$ seconds (s), and the average power consumption was 107.8 W. The second most energy-saving is ONNX Runtime, with an average of 107.3 W and a total execution time of $167.69\pm 4.72$ (s). The last framework is PyTorch, which consumes on average 113.9 W. VGG-16 model results are shown in Figure~\ref{fig:vgg16}. In this case, we notice comparable time performance using TensorFlow Lite and ONNX, with the last framework being less power-demanding. Specifically, TensorFlow Lite processes 100 images in $305.42\pm 4.65$ (s), requiring a power average of 117.8 W, while the ONNX employs $395.97\pm 4.75$ (s) to execute the model using 112.8 W power average. PyTorch requires $613.92\pm 29.71$ (s) to process the model, taking an average of $117.56$ W. Figure~\ref{fig:sys} shows MobileNetV2 power results. As in the previous cases, TensorFlow Lite and ONNX Runtime process the dataset in a comparable time, $50.32\pm0.01$ and $60.37\pm 0.02$, respectively. The first consumes an average of 89.3 W while the latter, 96.4 W. PyTorch is the framework that takes more time $667.57\pm4.63$ and is more energy demanding, consuming an average of 110.8 W.

\begin{figure}
    \centering
\includegraphics[width=0.65\textwidth]{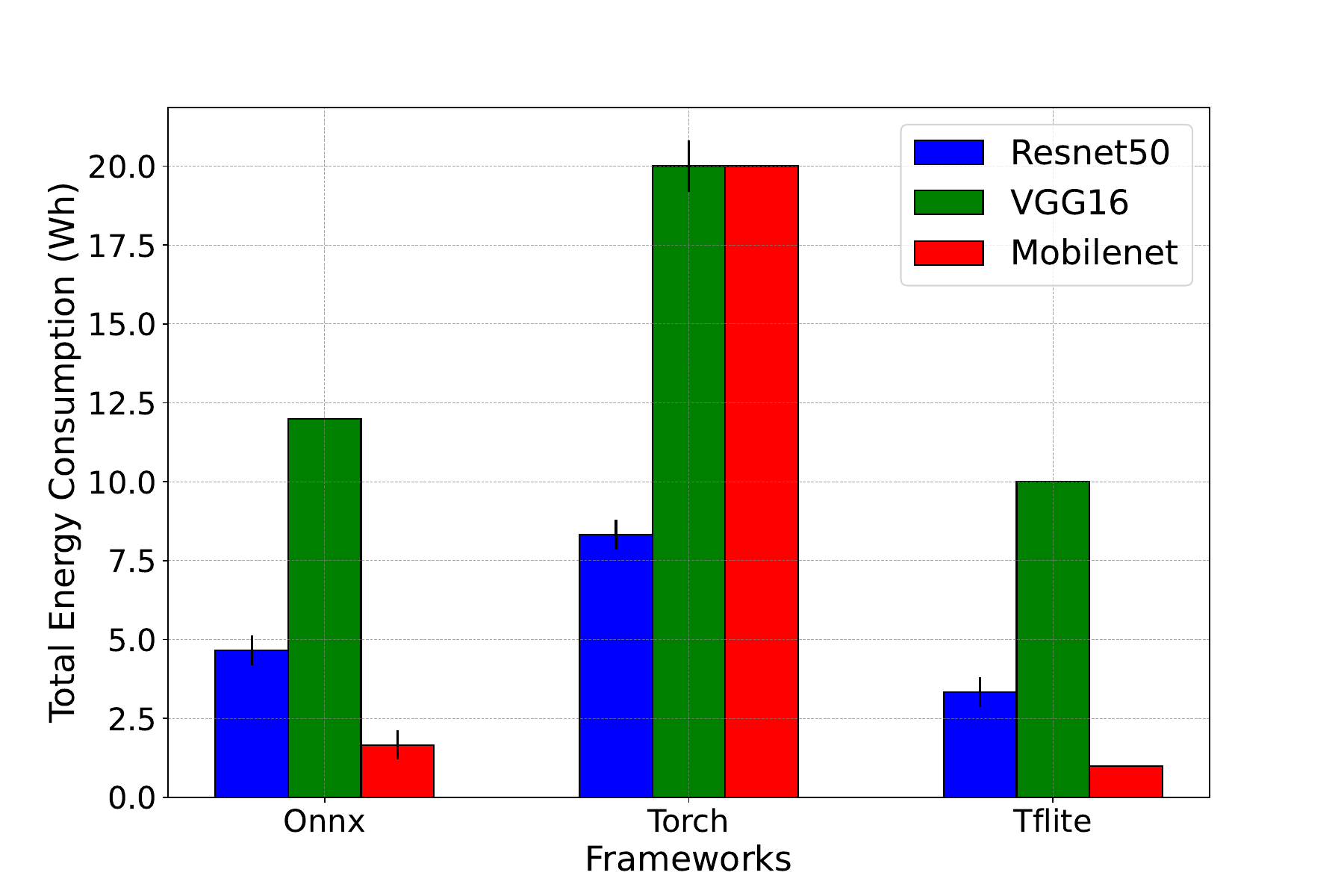}
    \caption{ResNet-50, VGG-16 and MobileNetV2 inference energy consumption, simulations on 1000 images.}
    \label{fig:energy}
\end{figure}

We also analyzed the energy consumed by the frameworks. Figure~\ref{fig:energy} shows the average energy of the models consumed by the frameworks. Consistently with the power consumption estimates, TensorFlow Lite is the most energy-efficient framework for the ResNet-50 model, consuming $3.33 \pm 0.47$ Wh. In comparison, ONNX Runtime and PyTorch are 1.39x and 2.42x more energy-intensive, respectively. Similar performance were obtained in the case of VGG-16, where ONNX Runtime and PyTorch consumed 1.2x and 2.0x more energy, respectively, compared to TensorFlow Lite. For the mobilenet model, PyTorch is the high energy-demanding framework, ~20x more than TensorFlow Lite. ONNX Runtime requires $1.66\pm 0.47$ Wh.

Concluding the analysis, we can say that TensorFlow Lite and ONNX Runtime have comparable performance in terms of both execution time and energy consumed. This is not surprising considering that both frameworks use XNNPACK to optimize inference. In contrast, PyTorch uses a native model implementation that is only optimized using OpenBLAS to speed up linear algebra operations. This may be the cause of the difference in performance. However, this requires a thorough investigation, which is beyond the scope of this paper.

\section{Conclusions}\label{sec6}
This work explores the power and energy consumption of inference model executions using three popular AI frameworks on RISC-V CPU processor. The results show that, in our cases, TensorFlow Lite and ONNX Runtime are the most performant and energy-saving frameworks, with the latter that is at maximum 1.6x less energy-demanding compared to the former. PyTorch shows slightly worse performance, in particular in the case of Mobilenet model. This could be due to different back-ends libraries utilized, XNNPACK vs native Torch+OpenBLAS. Future in-depth analysis may give further clarity on these differences in performance.

In the future, we plan to conduct a deeper investigation into the energy consumption of AI frameworks, examining various back-end acceleration libraries and considering different hardware architectures. Additionally, we aim to expand the number of frameworks included in our analysis.

\section*{Acknowledgements}
This work has been partially supported by the Spoke 1 ``FutureHPC \& BigData'' of the ICSC--Centro Nazionale di Ricerca in High Performance Computing, Big Data and Quantum Computing and hosting entity, funded by European Union—Next GenerationEU; and by the European Union under the project DYMAN (grant n. 101161930). 

\bibliographystyle{splncs04}
\bibliography{bibliography.bib}

\end{document}